\documentclass[aps,prl,twocolumn,nopacs,floats,superscriptaddress]{revtex4-1}
\usepackage{graphicx}
\usepackage{dcolumn}
\usepackage{bm}
\usepackage{amsmath}
\usepackage{color}

\definecolor{cardinal}{rgb}{0.6,0,0}
\definecolor{darkgreen}{rgb}{0,0.4,0}
\definecolor{golden}{rgb}{0.92, 0.7, 0}
\definecolor{midnight}{rgb}{0, 0, 0.5}
\definecolor{darkblue}{rgb}{0, 0, 0.7}

\def\he4{$^4$He}
\def\hel3{$^3$He}
\def\Am3{\AA$^{-3}$}
\def\beq{\begin{equation}}
\def\eeq{\end{equation}}

\newcommand{\pt}{{\partial}}

\newcommand{\oh}{{\frac{1}{2}}}

\newcommand{\cH}{{\mathcal H}}
\newcommand{\cL}{{\mathcal L}}

\newcommand{\be}{\begin{equation}}
\newcommand{\ee}{\end{equation}}
\newcommand{\bea}{\begin{eqnarray}}
\newcommand{\eea}{\end{eqnarray}}
\newcommand{\bse}{\begin{subequations}}
\newcommand{\ese}{\end{subequations}}
\def\rf#1{(\ref{#1})}

\begin{document}
%%%%%%%%%%%%%%%%%%%%%%%%%%%%%%%%%%%%%% AUTHORS %%%%%%%%%%%%%%%%%%%%%%%%%

\author{Leo Radzihovsky}
\affiliation{ Department of Physics and Center for Theory of Quantum Matter, University of Colorado, Boulder, CO 80309}

\author{Anatoly Kuklov}
\affiliation{Department of Physics \& Astronomy, College of Staten Island and the Graduate Center of
CUNY, Staten Island, NY 10314}

\author{Nikolay Prokof'ev}
\affiliation{Department of Physics, University of Massachusetts, Amherst, MA 01003, USA}

\author{Boris Svistunov}
\affiliation{Department of Physics, University of Massachusetts, Amherst, MA 01003, USA}
\affiliation{Wilczek Quantum Center, School of Physics and Astronomy and T. D. Lee Institute, Shanghai Jiao Tong University, Shanghai 200240, China}

%%%%%%%%%%%%%%%%%%%%%%%%%%%%%%%%%%%%%%%%%%%%%%%%%%%%%%%%%%%%%%%%%%%%%%%%%%%%%%
\title{Superfluid Edge Dislocation: Transverse Quantum Fluid}

%%%%%%%%%%%%%%%%%%%%%%%%%%%%%%%%%%%%%%%%%%%%%%%%%%%%%%%%%%%%%%%%%%%%%%%%%%%%%%
\begin{abstract}
  Recently, it has been argued [Kuklov {\it et al.},
  Phys. Rev. Lett. {\bf 128}, 255301 (2022)] that unusual features
  associated with the superflow-through-solid effect observed in solid
  \he4 can be explained by unique properties of dilute distribution of
  superfluid edge dislocations. We demonstrate that stability of
  supercurrents controlled by quantum phase slips (instantons), and
  other exotic infrared properties of the superfluid dislocations
  readily follow from a one-dimensional quantum liquid distinguished
  by an effectively infinite compressibility (in the absence of
  Peierls potential) associated with the edge dislocation's ability to
  climb.  This establishes a new class of quasi-one-dimensional
  superfluid states that remain stable and long-range ordered despite
  their low dimensionality. We propose an experiment to test our mass-current--pressure characteristic prediction.
%\\
%\noindent Keywords: Solid \he4; Superflow; Dislocations; Phase slip; Plasticity
\end{abstract}
%\pacs{ 67.80.bd, 67.80.dj, 67.80.-s, 67.80.B-}
% 67.80.bd Superfluidity in solid 4He, supersolid 4He
% 67.80.dj Defects, impurities, and diffusion
% 67.80.-s Quantum solids
% 67.80.B- Solid 4He

\maketitle

{\it Introduction.} About a decade ago the superflow-through-solid
(STS) effect in a structurally imperfect crystal of \he4
\cite{Hallock,Hallock2012,Hallock2019,Beamish,Moses,Moses2019,Moses2020,Moses2021},
along with the striking companion effect of anomalous isochoric
compressibility (also known as the syringe effect) \cite{Hallock} have
been attributed---by means of {\it ab initio} simulations---to the
properties of superfluid edge dislocations (SED) \cite{sclimb}.
However, two apparently unrelated experimental features, namely (i) an
exponentially strong suppression of the flow by a moderate increase in
pressure and (ii) an enigmatic temperature dependence of the flow
rate, remained unexplained until very recently, when both dependencies
were argued to be linked and accounted for by the highly unusual
properties of isolated SED \cite{Kuklov2022}. The arguments of
Ref.~\cite{Kuklov2022} rest on the self-consistent assumption that a
SED can support stable supercurrents despite being
quasi-one-dimensional and featuring the spectrum of elementary
excitations in violation of the Landau criterion (see the Supplemental
Material \cite{SM} for the discussion of the Landau criterion in a
non-Galilean superfluid system).

An outstanding (apparent) inconsistency between the scenario of
Ref.~\cite{Kuklov2022} and the experimental data has been the observation of mass-current--pressure (I-V) characteristic reminiscent of a Luttinger
liquid (LL) \cite{Hallock2012}, while the temperature dependence of
the flux was incompatible with the LL physics.

\begin{figure}[!htb]
%\vskip-8mm
	\includegraphics[width=1.0 \columnwidth]{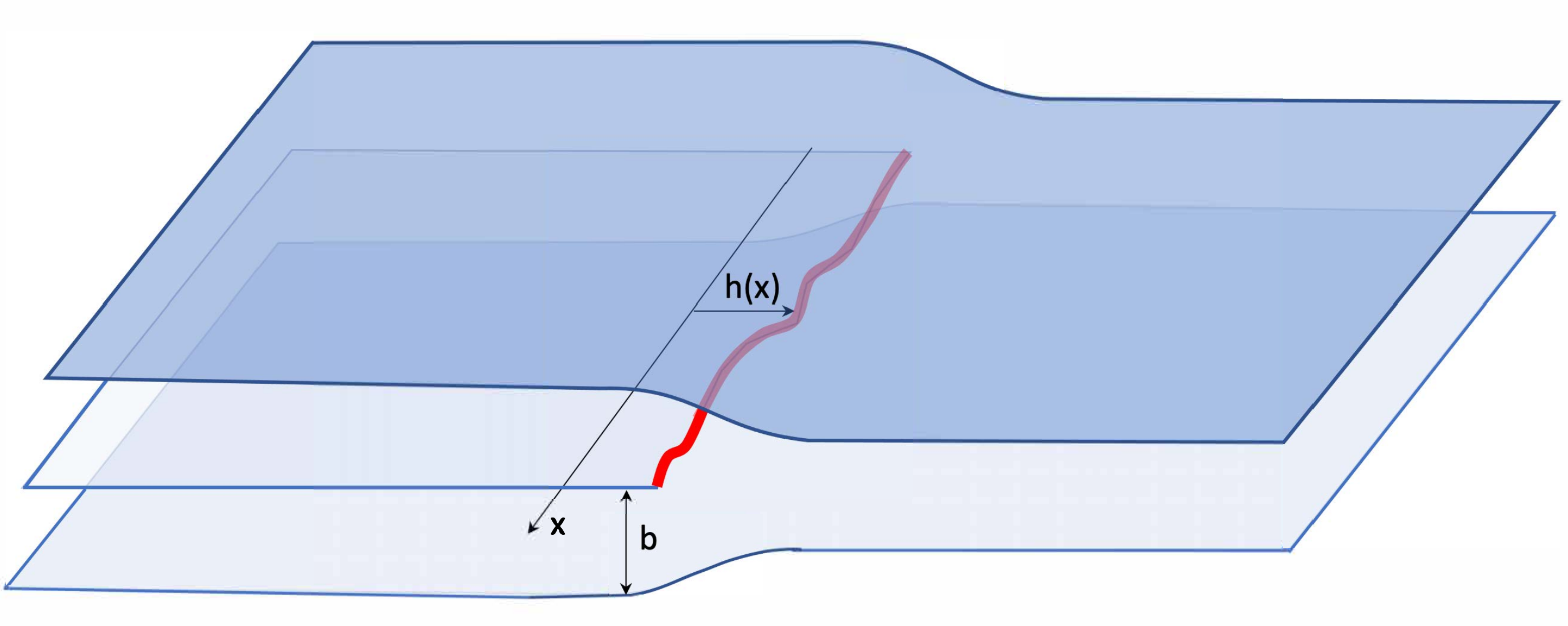}
%	\vskip-8mm
	\caption{(Color online) Sketch of the superfluid edge
          dislocation marked by the bold (red) wavy line. Its Burgers
          vector $ \bf b$ (along the C-axis) is perpendicular to the planes of
          atoms---two complete layers are shown as two parallel planes
          with straight edges and the incomplete one is limited by the
          dislocation line. The transverse translation invariance of
          $h(x)$ (continuous in the absence of the pinning Peierls
          potential, and discrete otherwise) is responsible for
          infinite compressibility of the bosonic fluid confined to
          the dislocation core.}
	\label{fig:SED}
\end{figure}

In this Letter, we put the theory of SED on a solid theoretical basis
by examining the consequences of the key feature distinguishing a SED
from a LL: SED is characterized by a (nearly) divergent linear
compressibility
%originating from
%(an approximate---see below)
imposed by the approximate
%(neglecting Peierls barrier, that reduces it to a discrete one)
translational invariance of the dislocation transverse to its Burgers
vector and its axis, as illustrated in Fig.~\ref{fig:SED}. The I-V
characteristic that we find resolves the above-mentioned inconsistency
with the experimental data and suggests a simple experiment to confirm
our theory.

{\it Transverse Quantum Fluid model.} The minimal model---which we
refer to as Transverse Quantum Fluid (TQF)---proposed in
Ref.~\cite{sclimb} for SED in an ideal crystal of \he4 with an average
superflow velocity $\sim v_0$ along its $x$-directed core, is given by
a 1D bosonic Hamiltonian, $H[\phi,n] = \int\cH \, dx $, with 
\be \cH =
{1\over 2\kappa} (\pt_x n)^2 + {n_s\over 2} (v_0 + \pt_x \phi )^2 \, .
\label{H}
\ee 
The 1D boson density $n$ is canonically conjugate to the
superfluid phase $\phi$, and the second term is the kinetic energy of
the flow, with the superfluid stiffness, $n_s$, exponentially sensitive
to the local pressure. The parameter $\kappa$ is determined by the inverse shear
modulus of the crystal \cite{sclimb}.  As shown in Ref.~\cite{Kuklov2022}, namely this
feature accounts for the unusual dependence of the critical current on
temperature.

The key feature distinguishing TQF from LL is the absence of the
compressibility term $\chi^{-1} n^2$, namely, TQF is characterized by
a divergent compressibility $\chi$.  The condition $\chi^{-1} = 0$ is
enforced by the translation invariance of the dislocation motion transverse
to its core axis and its Burgers vector (Peierls potential effects
that violate this symmetry and lead to finite compressibility are
discussed below). This feature is illustrated in Fig.~\ref{fig:SED}:
The dislocation climb displacement $\delta h(x)$ corresponds to
the 1D boson density change,
\be
\delta n(x)=\delta h(x)/a^2,
\label{nh}
\ee
along the dislocation, where $a$ is a lattice constant that we set to
unity hereafter.  The associated spectrum $\omega_k$ of elementary
excitations (at $v_0=0$) is straightforwardly found to be quadratic in
the wavevector $k$ along the dislocation,
\be \omega_k = D k^2 , \qquad D =\sqrt{n_s/\kappa},
\label{quadratic}
\ee
as a direct consequence of the aforementioned translation invariance,
that leads to $n\to n+{\rm const}$ invariance of (\ref{H}).

We note that the {\em quadratic} dispersion (\ref{quadratic}) appears
to violate the Landau criterion for the critical velocity,
$v_0 < \{ \omega_k /k \}_{\rm min}$, for arbitrary small $v_0$.
However, as we discuss in the Supplemental Material (SM) \cite{SM}, the Landau
criterion for instability relies on the existence of the term
$\delta n(v_0 + \pt_x \phi )^2 = 2 v_0\delta n \pt_x \phi +\dots$,
which is forbidden for SED by the climb translation invariance,
allowing only density derivatives, $\partial_x n$, in the TQF
Hamiltonian.  Thus, at sufficiently small $v_0$, SED (described by
TQF) does not develop the Landau instability, and we must consider
superflow relaxation via quantum phase slips, i.e., instantons
(space-time vortices) in the phase field $\phi(x,\tau)$ configuration
\cite{ColemanInstantons}.

The Euclidean Lagrangian density of TQF, corresponding to (\ref{H})
can be obtained after eliminating the density $n$,
\be
\cL   = {\kappa\over 2} (\pt_x^{-1}\pt_\tau \phi)^2
+ {n_s\over 2} (\pt_x \phi )^2  + n_s v_0  \pt_x \phi  ,
\label{L2}
\ee
where the long-range operator $\pt_x^{-1}$ is defined
by its Fourier transform, namely, $\pt_x^{-1} \to  -i k^{-1}$.
The last (boundary) term in (\ref{L2}) is important only in the presence of
instantons, and we have omitted an irrelevant constant.

{\it Goldstone modes and long-range order.} Mean-squared fluctuations
of the superfluid phase $\phi$, computable within the Gaussian
approximation,
\begin{equation}
  \langle\phi^2\rangle
  =\frac{1}{\kappa}\int^{2\pi/a}\frac{d\omega dk}{(2\pi)^2}
  \frac{k^2}{\omega^2 +  D^2 k^4}
  \approx \frac{1}{a\sqrt{\kappa n_s}}\, ,
      \label{phiRMS}
\end{equation}
are finite, and TQF thus features a long-range order in the superfluid
field $\psi = e^{i \phi}$ at zero temperature ($T$), despite being
one-dimensional \cite{ODLRO}. Given the exact mathematical symmetry
between $\phi $ and $n$ in (\ref{H}), the field $\eta=e^{ i 2\pi n}$
also exhibits long-range order at zero-temperature.

At this point we recall that SED is a subject
to a Peierls potential, $U = U_0\cos (2\pi n)$, [utilizing the density
$n$ -- displacement $h$ relation, \rf{nh}], omitted in the formulation
(\ref{H}). This sets certain limitations on the applicability of the TQF as an asymptotic 
model for SED.  Long-range order in $\eta $ implies that $U$ is a relevant
perturbation, which suppresses the dislocation climb motion and
induces a crossover of the excitation spectrum  to a conventional (LL-type) linear  form 
 at small momenta, $k/(2\pi) < \xi^{-1}$, 
with the length scale $\xi$ diverging with crystal's vanishing shear modulus $\sim \kappa^{-1}$
\cite{MAX}. (In the opposite limit, the LL superfluid undergoes a
phase transition to a Mott-Insulator \cite{MAX}.) Concomitantly, at
length scales longer than $\xi$, the zero-temperature long-range order
of SED changes to the algebraic quasi-long-range one, typical for LL.
Such a crossover was observed in model simulations of
Refs.~\cite{MAX,ODLRO} through a finite-size scaling of
compressibility $\chi$ for a pinned dislocation with $h(0)=h(L)=0$:
from TQF's $\chi \sim L^2$ to a constant in the $L\to \infty$ limit, as
$T \sim 1/L \to 0$.

In what follows, we will focus on the exotic TQF regime on scales
below $\xi$, where SED is quantum rough and the associated 1D
superfluid is long-range ordered. 
 Equivalently, one can view TQF as the asymptotic limit ($\xi \to \infty$) of  SED. We demonstrate that such a simplification is
sufficient for explaining the experimental data with high accuracy.
%For a weak Peierls potential, particularly with a Burgers vector along a low symmetry axis, this novel regime appears at extensive {\color{green} length scales shorter than the crossover scale to the asymptotic LL state and will be the one of experimental relevance in $^4$He.}

{\it Confinement of instantons.} We now analyze the stability of this
novel one-dimensional superfluidity to superflow-induced instantons
\cite{ColemanInstantons}. To account for these, we consider
non-single-valued configurations of the superfluid phase
$\phi(x,\tau)$, corresponding to its space-time vortex
configurations. To this end, we introduce the velocity field
$v_\mu \equiv \pt_\mu\phi$ in the $(1+1)$-dimensional space-time
$x_\mu = (x,\tau)$. The instantons have the form of point-vortex
singularities in the otherwise regular field $v_\mu$:
\begin{equation}
  \pt \times v  = q(x_\mu)
  = \sum_{j} \,  q_{j} \, \delta^2 (x_\mu - x_{\mu,j}) ,
\label{instanton}
\end{equation}
where $\pt \times v \equiv\epsilon_{\mu\nu}\pt_\mu v_\nu$ is
short-hand notation for $(1+1)$ space-time curl of $v_\mu$ and $q_{j}$
and $x_{\mu,j}$ are, respectively, the ``charge" (an integer multiple of
$2\pi$) and space-time position of the $j$-th instanton.

With these ingredients we can now straightforwardly derive the
instanton action $S[q(x_\mu)]$ from the Lagrangian density \rf{L2} for
$\phi(x_\mu)$.  The most direct way to obtain this (with
alternative derivations presented in Ref.~\cite{SM}) is to
enforce the topological constraint \rf{instanton} using a functional
delta function in a path-integral for a partition function over
$v_\mu$,
\begin{equation}
  Z = \int [dv_\mu][dq][d\lambda] e^{-\int dx d\tau \cL[v_\mu,\lambda,q]} \, ,
\label{Z}
\end{equation}
with the Lagrangian density (for $v_0=0$)
\begin{eqnarray}
  \cL   &=& \frac{\kappa}{2} (\pt_x^{-1}v_\tau)^2 +
            \frac{n_s}{2}(v_x)^2 + i\lambda(\pt \times v - q)\, .
         \label{LL1}
\end{eqnarray}
In $\cL$ we implemented the vorticity constraint \rf{instanton} via a
functional delta function as an integral over the auxiliary field
$\lambda(x_\mu)$.  Performing the Gaussian integration over fields
$v_\mu(x_\mu)$ and $\lambda(x_\mu)$, gives the instanton action,
\bea
S &=&  { n_s \over 2} \int {d\omega d k\over (2\pi)^2} \, { | q_{\omega,k}|^2 \over \omega^2 + D^2 k^4 } \\
&=& \oh\int_{x_\mu, x'_\mu} q(x_\mu)V(x_\mu - x'_\mu) q(x'_\mu)\; ,
\label{interactionS}
\eea
where $V(x, \tau)$ is the space-time instanton-instanton
interaction, which after subtracting the self action $V(0,0)$, becomes
\begin{eqnarray}
V(x, \tau)
&=& n_s \int\frac{d\omega dk}{(2\pi)^2}\,
    \frac{e^{i (\omega \tau + k x )}-1}{\omega^2 +  D^2 k^4} \nonumber\\
 &\approx&
 \left\{\begin{array}{ll}
-\frac{n_s \sqrt{|\tau|}}{2\sqrt{\pi D}},\;\; \mbox{ for $ x^2 \ll
          D\tau$},\\
- \;\frac{n_s|x|}{4D}, \;\;\;\;\;\mbox{ for $ x^2 \gg D\tau$.}
        \end{array}\right.
  \label{Vinteraction}
\end{eqnarray}
This result can be complementarily obtained (see Ref.~\cite{SM}) through the saddle-point
equation for $v_\mu$ together with the Fourier-transformed instanton
constraint, Eq.~(\ref{instanton}), finding
\be
  v_\tau \, =\, {- i D^2 k^3  q \over D^2  k^4 +  \omega^2 }\, ,
  \qquad  v_x \, =\, {i \omega q \over D^2 k^4 + \omega^2 }  \, .
\label{velocities}
\ee

The interaction kernel in \rf{Vinteraction} is reminiscent (but misses
a factor of $-\pt_x^2$ in the numerator) of interaction between
dislocations in the 2D classical smectic \cite{TonerNelson81}. This
difference is expected as the 2D smectic elasticity is ``softer'' than
in the XY model, while here it is ``stiffer'' than in the XY model
because of the extra $\pt_x^{-1}$ factor in $\cL$ of \rf{L2}.  The
latter effect is directly related to divergent compressibility, which
allows stronger density fluctuations and thereby ``stiffens'' the
canonically conjugate superfluid phase $\phi$.  As a result, in
contrast to the 2D smectic (where interaction is weak and dislocations
are deconfined) \cite{TonerNelson81,QSmGaugeLR}, the TQF instantons
with opposite ``charges" are {\it always confined} by the interaction
$V(\tau,x)$ in \rf{Vinteraction}, featuring a power-law growth with
spatial and temporal separation.  Note that this behavior is
consistent with the 1D long-range superfluid order at $T=0$ discussed
above \cite{ODLRO}.

%%%%%%%%%%%%%%%%%%%%%%%%% HERE %%%%%%%%%%%%%%%%%%%%%%%%%%%%%%%%%%%%%

{\it Metastability of the superflow and nonlinear I-V response.} We
now demonstrate that instanton confinement, \rf{Vinteraction}, leads
to the exponential in $1/v_0$ metastability of the superflow. This is
readily seen by examining the contribution of the $v_0$-dependent term
to the Langragian density \rf{L2}.  Since this is a boundary term, its
presence does not change the conditions \rf{instanton} and action in
\rf{Vinteraction}, leaving the solution \rf{velocities} intact.
Further simplification comes from the fact that the metastable regime
corresponds to appropriately small values of $v_0$, when the
destabilization channel is associated with a well-isolated
instanton--anti-instanton pair characterized by space-time coordinates
$(x_+,\tau_+)$ and $(x_-,\tau_-)$.  Straightforward integration 
in the $v_0$-dependent term in \rf{L2}, $ \int dx \, \partial_x \phi (x,\tau)= \phi(+\infty, \tau ) - \phi(-\infty,\tau)$, gives its
contribution to the instanton pair action (for the pair with ``charges" $\pm 2\pi$)
\bea
S_{v_0} \, =\,  v_0 n_s \iint d\tau dx  v_x \, &=&\,  2\pi v_0 n_s\int_{\tau_+}^{\tau_-} d\tau \nonumber \\
&= &\, 2\pi v_0 n_s ( \tau_+ - \tau_-) \, . \qquad
\label{pair_v_0}
\eea
We see that, in a close similarity with the vortex--anti-vortex pair
in a 2D superfluid, the superflow generates a transverse (in
space-time) ``force'' pulling the pair apart along the
imaginary-time direction.

Combining this with the competing $v_0$-independent part of the pair
action at short distances, $S_{\rm inst}^{(0)}\approx V(x = 0, \tau)$,
given by \rf{interactionS}--\rf{Vinteraction}, we find the total
instanton-pair action,
%
%\be
%S_{\rm inst}^{(0)} \approx V(x = 0, \tau) \approx
%\frac{n_s}{\sqrt{D}} \sqrt{|\tau|}\; ,
%\label{instantonPairS}
%\ee
%
\begin{eqnarray}
S_{\rm inst}
&\approx&  \frac{2\pi^{3/2}n_s}{\sqrt{ D}} \sqrt{|\tau|} -2\pi n_s|v_0\tau|  \, ,
\label{instantonPairSscaling}
\end{eqnarray}
where
$(x,\tau)\equiv (x_+ - x_-, \tau_+ - \tau_-)$.  The maximum-action
instanton-pair configuration controlling the dissipation is reached at
$\sqrt{\tau^*(v_0)} = \sqrt{\pi} /(2\sqrt{D} v_0) $, and the
corresponding action is given by
\be S_{\rm inst}^* \, \equiv \, S_{\rm
  inst}(\tau^*) \, \approx\, \frac{ \pi^2 n_s }{2 D v_0}\, .
\label{instantonPairOpt}
\ee
In the exponential approximation \cite{ColemanInstantons}, this defines
the probability of the instanton nucleation:
\begin{equation}
  {\cal P} \sim  e^{-v_c/v_0}, \qquad
  v_c = \frac{\pi^2 n_s }{2 D}= \frac{ \pi^2 \sqrt{n_s\kappa}}{2},
  \label{notIV}
\end{equation}
in sharp contrast with the LL's power-law dependence
${\cal P} \sim (v_0)^g$, with $g>0$.

Another qualitative difference with the LL physics comes from
kinematic considerations regarding the ultimate decay of the superflow.
In the translation-invariant LL, the decay of the supercurrent is
kinematically forbidden because it is impossible to simultaneously
satisfy the conservation of energy and momentum under generic
conditions. One needs either impurities and/or commensurate external
potential to absorb the momentum released by the supercurrent when the
phase winding number changes by $\pm 1 $ due to a quantum phase
slip.  In TQF, the decay of the supercurrent into elementary
excitations is kinematically allowed due to the quadratic dispersion,
Eq.~\rf{quadratic}.  However, at small values of $v_0$, this
involves a large number of elementary excitations.
 Indeed, the energy
and momentum released by a supercurrent in a single phase slip are
(respectively): $\Delta E = 2\pi n_s v_0$ and $\Delta P = 2\pi n_s$.
Suppose these are absorbed by ${\cal N}$ quasiparticles with
momentum $k_*=\Delta P /{\cal N}$. From the energy conservation,
$\Delta E = {\cal N} D \Delta P^2/{\cal N}^2 = D \Delta P^2/{\cal N} $, we then readily
obtain ${\cal N} = D \Delta P^2/\Delta E  = 2\pi n_s D /v_0$ and
$k_* = v_0/D$. In what follows we will not be pursuing further the analysis of 
specific details of the decay and assume that the exponential factor in Eq.\rf{notIV} controls
the rate of the phase slips.

    %The IV characteristic of the supercritical flow along the dislocation.}
  \begin{figure}[!htb]
%\vskip-8mm
	\includegraphics[width=0.9 \columnwidth]{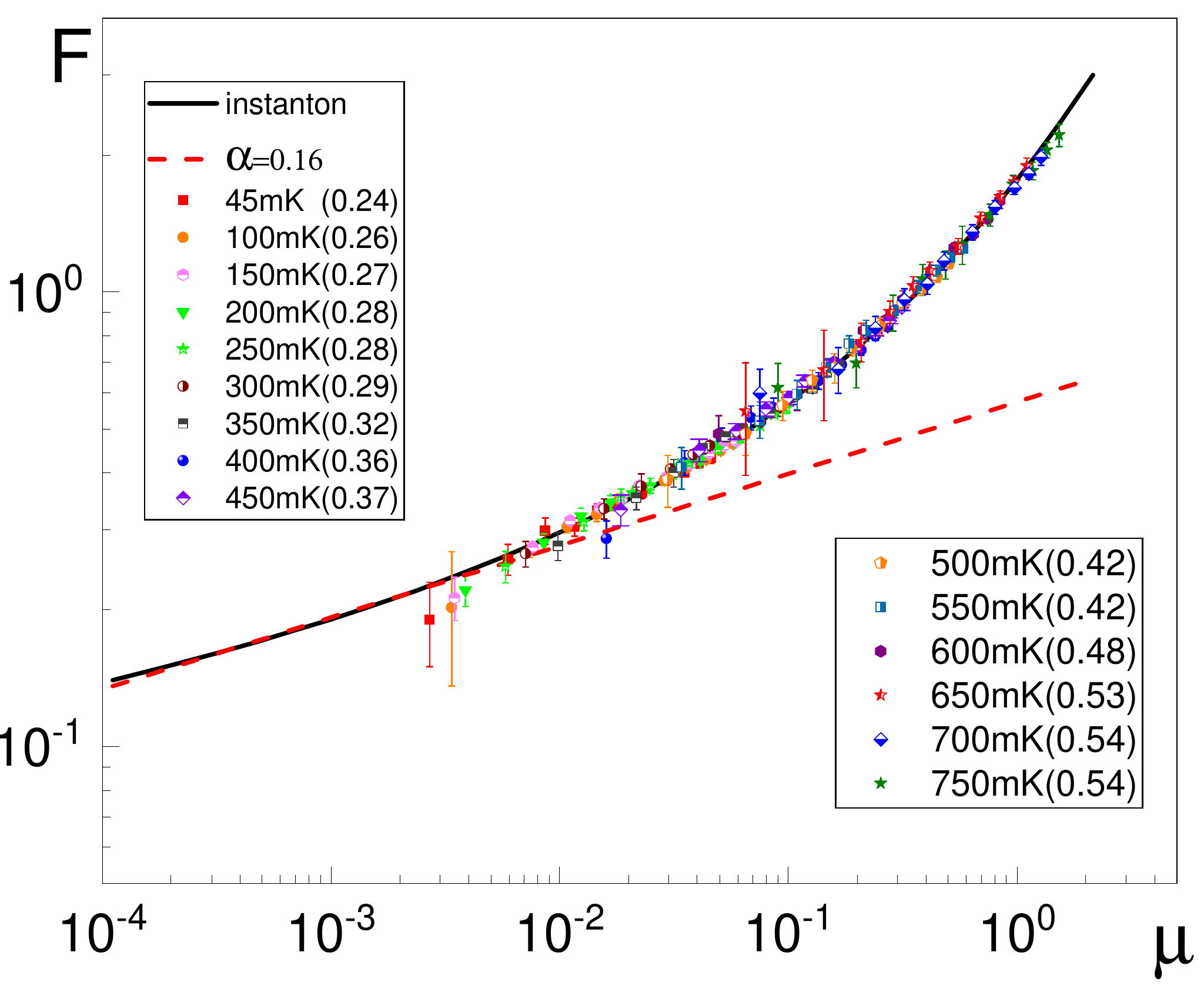}
%	\vskip-8mm
	\caption{(Color online) The I-V characteristic master curve (solid line) of the TQF  ,
          Eq.~(\ref{vmu})  (units are arbitrary).  Symbols are the experimental
          data from Fig.~5(a) of Ref.~\cite{Moses2019} [presented also in Fig.1 of Ref.~\cite{SM}]  collected
          at different $T$ and interpreted  within the sub-Ohmic dependence (\ref{IVLL}).  
The corresponding values of $T$ and $\alpha$ (in parenthesis) are shown in the legend.
          The
          data points from a set at a given $T$ are shifted  without changing their log-log
          slopes to achieve the collapse  onto the master curve (see the text below).%rescaled          vertically and horizontally to achieve a better collapse
          The dashed line, as an example, corresponds to the power law
          (\ref{IVLL}) with $\alpha =0.16$.}
	\label{fig:IV}
\end{figure}

{\it Experimental implications.} The results obtained above allow us
to resolve an apparent disagreement between the scenario advocated in
Ref.~\cite{Kuklov2022} and the experimental data on the I-V
characteristic of the STS effect.  Experimentally, the flow
rate $F$ (the mass current, proportional to $v_0$) as a function of
the chemical potential bias (the ``voltage") $\Delta \mu$ was found to
be consistent with the sublinear power-law
\cite{Hallock2012,Moses2019}:
\be F  = A(T) (\Delta \mu)^\alpha, \qquad  ( \alpha<1) .
\label{IVLL}
\ee
The authors of Ref.~\cite{Hallock2012} reported
$\alpha\approx 0.3\pm 0.1$, independent of $T$ up to $T\approx 0.5$K.
This value agrees well with $\alpha \approx 0.24$, observed at
$T<0.2$K in Ref.~\cite{Moses2019}.  However, as $T \to 1$K, the value
of $\alpha(T)$ was found to cross over to $\alpha\approx 0.5$
\cite{Moses2019}.  If it were not for the drastic temperature
dependence of the flux amplitude, $A(T)$, as well as the less shocking
but still unexpected temperature dependence of $\alpha(T)$, it would be natural to interpret (\ref{IVLL}) as the
manifestation of the LL behavior \cite{Hallock2012}.

By suggesting a solution for the temperature dependence of the
amplitude $A(T)$, Ref.~\cite{Kuklov2022} questioned the LL origin of
the dependence (\ref{IVLL}), without providing a detailed
mechanism. Present analysis offers a concrete alternative to the LL
interpretation based on the TQF instanton mechanism of the phase slips
described above by Eqs.~(\ref{instantonPairOpt})--(\ref{notIV}).  If
the probability of phase slips is controlled by the TQF instanton
action (\ref{instantonPairOpt}) rather than by the matrix element of
the transition from the initial to the final state of the system, the
I-V characteristic of the superfluid edge dislocation can be obtained
within an elementary hydrodynamics approach (see Supplemental Material
\cite{SM}).  Accordingly, the resulting I-V curve  can be described by
the relation 
%\be
${\cal P}_0 v_0 {\cal P} =\Delta \mu $ %{\rm e}^{-v_c/v_0}= \Delta \mu$,
%\label{fIV}
%\ee
where ${\cal P}$ is given in Eq.\ref{notIV} and ${\cal P}_0$ is a constant that includes the details of the
mechanisms for the transfer of the total momentum and energy from the
current to the excitations.  Then, introducing dimensionless variables
$\hat v=v_0/v_c, \,\, \mu = \Delta \mu/({\cal P}_0 v_c)$, the
normalized flow velocity $\hat v$ (in units of critical velocity,
$v_c$),
%and the flow rate through one dislocation, $n_s v_c \hat v$,
obey the relation
\be \hat v \, {\rm e}^{-1/\hat v}=\mu \, ,
\label{vmu}
\ee
with the flux $F$ through a sample containing many dislocations being $F\propto \hat{v}$. 
While the overall curve $\hat v$ vs $\mu$ is obviously inconsistent
with the dependence (\ref{IVLL}), its significant parts can be well
fit by this dependence, provided the interval of $\mu$ variation does
not exceed two orders of magnitude -- as demonstrated by the dashed line in Fig.~\ref{fig:IV}. %{\color{red} This is illustrated by the dashed line $F\sim \mu^\alpha, \,\, \alpha=0.16,$ in Fig.~\ref{fig:IV}.As it turns out, the bias by chemical potential in the experimental data for any given temperature $T$ \cite{Hallock2012,Moses2019} (see also Fig.1 in Ref.~\cite{SM}) is restricted by two orders of magnitude. However, as explained below, the effective range of the bias depends on $T$, and thus covers about four orders of magnitude, which immediately reveals significant deviations from the dependence (\ref{IVLL}).}

  In accordance with the
large-fluctuation scenario of Ref.~\cite{Kuklov2022}, the crossover
between different parts of the curve takes place with increasing
temperature when at fixed $\Delta \mu$, the dimensionless bias
$\mu \sim \Delta \mu/\sqrt{n_s(T)}$ increases exponentially, shifting the I-V
characteristic from $\alpha\approx$ 0.24 at low $T$ towards
${\hat v} \sim 1$, where $\alpha\approx 0.5\div 0.6$. This behavior is
demonstrated by the good collapse of the data $F$ vs $\mu$  \cite{Moses2019} on the master curve in
Fig.~\ref{vmu}. [See more details in Ref.~\cite{SM}]. %, and it is in line with the experimental observations \cite{Moses2019}. 
%Furthermore, as shown in the SM \cite{SM}, the $T$-dependencies of the rescaling factors $F_0(T), \mu_0(T)$ of the data points in $F(\Delta \mu, T)\to F_0(T) F(\Delta \mu, 0)$ and $\Delta \mu \to \mu_0(T) \Delta \mu $ (used to collapse them on the master curve) are consistent with the temperature dependencies described in Ref.~\cite{Kuklov2022}.
%(or, equivalently, $v_0$ does not change by a factor larger than $\sim 3\div 5$).
%{\color{red} The collapse procedure consists of the rescaling $F(\Delta \mu, T)\to F_0(T) F(\Delta \mu, 0)$ and $\Delta \mu \to \mu_0(T) \Delta \mu$ with the parameters $F_0(T), \mu_0(T)$ chosen so that the original data $F$ vs $\Delta \mu$ for each given $T$ can be shifted without changing its slope on the log-log scale. .} 

{\it Summary and outlook.} Motivated by a recent suggestion that phase
slips in a superfluid edge dislocation are expected to be
qualitatively distinct from those in a LL and allow for a metastable
superflow in 1D, we presented and analyzed the TQF model as the natural asymptotic limit of SED. 
 We demonstrated that  quantum Lagrangian of TQF
predicts confinement of quantum phase slips, implying the {\em
  exponential} nonlinear I-V characteristic,  along with other special
properties that have no analogs in known one-dimensional
systems.

This confinement of instantons, accompanied by the quadratic
dispersion of elementary excitations---the hallmark of the
TQF---follows directly from its divergent compressibility in the
absence of the Peierls potential. %Moreover, the ground state of the TQF features a genuine long-range off-diagonal order at zero temperature (a Bose-Einstein condensate), in spite of its one-dimensionality. 
Despite superficial similarity with the ideal Bose gas, the physics here is qualitatively distinct because (i) SED lacks Galilean invariance and (ii) translational invariance in the climb direction prohibits terms in the TQF Hamiltonian that are responsible for the Landau-type instability---a vanishing critical current -- for a parabolic dispersion in the ideal Bose gas.

Our  theory is consistent with the existing experimental
data on the superflow-through-solid effect thus offering a resolution of a vexing controversy in the data interpretation.
The exponential I-V characteristic that we predict is in a stark
contrast with the LL power-law, that requires a variable exponent to
fit the experimental data.  This naturally brings up a proposal for
the compelling experimental test: detailed measurements of the I-V
characteristic at low temperature and different external pressures and
then collapsing the data on the master curve predicted by Eq.~\rf{vmu}
and illustrated in Fig.~\ref{fig:IV}. Specifically, (i) extending the range of the biases
$\Delta \mu$ should demonstrate the deviation from the LL dependence (\ref{IVLL}),
(ii) applying external pressure at small $T$, should decrease $n_s$ and, accordingly, shift the effective $\alpha$ from 0.24 to higher values,
(iii) decreasing temperature below 45mK (the lowest studied in Ref.~\cite{Moses2019}) should either decrease $\alpha$  to even smaller
values---as demonstrated by the dashed line in Fig.~\ref{fig:IV}---or eventually bring the dislocation to the 
LL regime, where $\alpha$ is determined by the emerging finite compressibility.

{\it Acknowledgents}.  This work was supported by the National Science
Foundation under the grants DMR-2032136 and DMR-2032077.  LR
acknowledges support by the Simons Investigator Award from the Simons
Foundation. We thank Robert Hallock and Moses Chan for providing us with their data and for discussion of our results.

\vspace{1cm}
{\Large \bf Supplemental Material}

\section{Instanton Action: Extremal Configuration of the Velocity Field}

Given that TQF action  $S =\int {\cal L}  \, dx d\tau $  is bi-linear with respect to $\phi$ (apart from the linear in $\phi$ term for non-zero $v_0$) and that the dependence on $\phi$ is through the velocity components: $S[\phi] \equiv S[v_x, v_\tau]$ [see Eq.~(8) of the main text], the problem of the instanton action can be reduced to the problem of finding the velocity field minimizing the action at given instanton positions. In this section, we use a straightforward approach to the problem. In the next two sections, we will present several complementary analyses.

Let $\delta \varphi$ be a small single-valued variation of $\phi$ and $B[\phi]\equiv B[v_x, v_\tau]$
 the variational derivative of the action with respect to $\delta \varphi$:
\[
\delta S[\phi] \, =\, \int dx d\tau B[v_x, v_\tau] \, \delta \varphi  \, ,
\]
\[
B[v_x, v_\tau]\, =\,   G^{-1} \partial_x^{-2} \partial_\tau v_\tau - n_s \partial_x v_x \, .
\]
For the extremal configuration of the velocity field, we have $B[v_x, v_\tau]\equiv 0$ corresponding to
\be
\partial_\tau v_\tau - D^2 \partial_x^3 v_x \, =\, 0 \, .
\label{BB}
\ee
Combining (\ref{BB}) with the condition [Eq.~(6) of the main text]
\[
 \epsilon_{\mu\nu}\pt_\mu v_\nu  =  q \, ,
\]
we obtain a system of two equations relating the fields $v_x$ and  $v_\tau$ to $q=2\pi p, \,\, p=\pm 1, \pm 2, ...$.
Fourier-transforming these equations,
\be
 \omega v_\tau  \, +\,  D^2 k^3 v_x \, =\, 0\, ,
\label{BB_F}
\ee
\be
k v_\tau - \omega v_x \, =\, -i q \, .
\label{constraint_F}
\ee
This readily gives the Fourier transforms of $v_x$ and  $v_\tau$ as functions of the Fourier transforms of $q$, thereby arriving at Eq.~(12) of the main text.

\section{Euler-Lagrange equation in a generic gauge}

Here we utilize the fact that the field $\phi$ minimizing the action at fixed positions of instantons can be represented---in a continuum of different ways---as the sum of the smooth single-valued part $\phi_s$
and the part $\phi_0$ accounting for the instantons:
\be
\phi  = \phi_s + \phi_0 , \qquad   \quad  \epsilon_{\mu\nu}\pt_\mu  \pt_\nu \phi_0  =  q ,
\label{phi_s_Phi_0}
\ee
with a particular choice of $\phi_0$ corresponding to a particular ``gauge."
By substituting this decomposition inside the Lagrangian, and minimizing it
over $\phi_s$, we obtain the Euler-Lagrange equation of the form
\be
( D^2 \pt_x^4 -\pt^2_\tau ) \phi_s = \pt_\tau v^{(0)}_\tau - D^2 \pt_x^3 v^{(0)}_x \, ,
\label{E_L_gauge}
\ee
where $v^{(0)}_\mu = \pt_\mu \phi_0 $, so that
\be
 \epsilon_{\mu\nu}\pt_\mu  v^{(0)}_\nu  =  q \, .
\label{cond_def}
\ee
Gauge freedom amounts to the choice of $v^{(0)}_\mu$ satisfying the condition (\ref{cond_def}).

There is a special gauge reproducing the theory of the previous section. In this gauge, we require that  the right-hand side
of (\ref{E_L_gauge}) be identically zero, which, on one hand, leads to Eq.~(\ref{BB})  for  $v^{(0)}_\mu$ and, on the other hand, implies the trivial solution $\phi_s \equiv 0$, meaning $v_\mu \equiv v^{(0)}_\mu$.

Alternatively, we can use a simple gauge that nullifies one of the $v^{(0)}_\mu$ components---for definiteness, let it be $v^{(0)}_x$:
\be
 v^{(0)}_x=0\, , \qquad \quad  v^{(0)}_\tau=\pt_x^{-1}q\, .
 \label{our_gauge}
 \ee
 Substitutinq this into the r.h.s. of (\ref{E_L_gauge}) we find $\phi_s$ (in the operator form):
\be
\phi_s = { \pt_x^{-1} \pt_\tau q \over D^2 \pt_x^4 -\pt^2_\tau} \, ,
\label{phi_s}
\ee
which then yields the expressions for the two velocity components,
\be
v_x = v^{(0)}_x+ \pt_x \phi_s = { \pt_\tau q \over D^2 \pt_x^4 -\pt^2_\tau} \, ,
\label{v_x_phi}
\ee
\be
v_\tau = v^{(0)}_\tau+ \pt_\tau \phi_s = {D^2 \pt_x^3 q \over D^2 \pt_x^4 -\pt^2_\tau} \, ,
\label{v_x_phi}
\ee
reproducing result (12) in the main text.

\section{Constrained Euler-Lagrange equations for components of the velocity field}

Here we observe that since the two components of the velocity field,
\be
v_x = \partial_x \phi , \qquad   v_\tau = \partial_\tau \phi ,
\label{v_x_v_tau}
\ee
are allowed to be singular and thereby have both longitudinal and transverse parts, they
can be conveniently treated as independent fields subject to the constraint
\be
\partial_x v_\tau  - \partial_\tau v_x  = q .
\label{constraint}
\ee
In contrast to the phase $\phi$, its spatial and temporal derivatives are single-valued functions (with isolated singular points corresponding to instantons),
and this is one of the advantages of parametrization (\ref{v_x_v_tau})--(\ref{constraint}). The role played by the constraint (\ref{constraint}) is two-fold. On the one hand,
it takes into account the fact that the two components are not independent, so that there is only one independent physical degree of freedom. On the other hand, it provides a simple
way of including the positions of instantons into the action in a ``gauge-invariant" form.

 In its Fourier representation, the constraint is given by Eq.~(\ref{constraint_F}), so that
 the Lagrangian of corresponding extended action reads
\be
{\cal L}'  = {\cal L} + [\lambda^* (k v_\tau - \omega v_x) + {\rm c.c.}] ,
\label{L_prime}
\ee
where
\be
{\cal L}  = {\kappa \over 2k^2} |v_\tau|^2 + {n_s\over 2}|v_x|^2,
\label{L}
\ee
is the unconstrained part of the Lagrangian and $\lambda$ is the Fourier transform of the Lagrange multiplier.

The Euler-Lagrange equations immediately allow us to relate each of the two velocity components to $\lambda$,
\be
v_\tau   =  - 2\kappa^{-1}k^3 \lambda  ,  \qquad  v_x  =  {2\omega \over n_s} \lambda  ,
\label{E_L}
\ee
that leads to
\be
{\cal L}  = {2\over  n_s} (D^2 k^4 + \omega^2)|\lambda|^2  .
\label{L_of_lambda}
\ee
Substituting $v_x$ and $v_\tau$ from (\ref{E_L}) into (\ref{constraint_F}) to express $\lambda$ in terms of $q$, we obtain
\be
\lambda =  {i\over 2}  { n_s q \over D^2 k^4 +  \omega^2} \,  
\label{lambda_tilde_n}
\ee
and find the Lagrangian (12) of the main text.

\section{Duality transformation}

Here we start with the action (3) of the main text for $v_0=0$,
and introduce the Hubbard-Stratonovich field $j_\mu = (n, j_x)$
in the density-phase $(n,\phi)$ path integral to obtain
\begin{eqnarray}
  \cL &=& i n \pt_\tau\phi + i j_x\pt_x\phi  + \oh \kappa^{-1} (\pt_x n)^2 + \oh
          n_s^{-1} j_x^2 , \qquad   \label{Ldua1l}\\
&=& i j_\mu v_\mu  + \oh \kappa^{-1} (\pt_x n)^2 + \oh
          n_s^{-1} j_x^2 .   \label{Ldua2l}
\end{eqnarray}
By integrating out the smooth field $\phi_s$, we arrive at the
charge-conservation constraint, $\pt_\mu j_\mu = 0$,
i.e.,
\begin{eqnarray}
\pt_\tau n + \pt_x j_x = 0\;.
  \label{cont}
\end{eqnarray}
This is solved by
$j_\mu = \epsilon_{\mu\nu}\pt_\nu\theta\equiv(\pt\times\theta)_\mu$,
with
\begin{equation}
  n =\pt_x\theta\; ,\qquad j_x = - \pt_\tau\theta\;.
\label{solveCont}
\end{equation}
By substituting this solution into the Lagrangian, integrating the first term
by parts, and using the definition $\pt\times j = q$, we find,
\begin{eqnarray}
  \cL &=&  i q \theta  + E_c q^2 + \frac{1}{2\kappa} (\pt^2_x\theta)^2 + \frac{1}{2n_s}
              (\pt_\tau\theta)^2 ,   \label{Ldua3l}
\end{eqnarray}
where to account for lattice scale physics in this continuum treatment
we added the instanton core energy, $E_c$.  For $E_c = 0$,
integration over $\theta$ reproduces our result for the instanton action (9),(10) of the main text.

\section{TQF as the asymptotic limit model for SED}
As explained in the main text, the TQF is the asymptotic model for SED -- that is,
at scales shorter than $\xi$.
 In the SED at scales longer than $\xi$, the pinning effects of the Peierls
potential change the denominator of the solution in Eq.(12) of the main text
from $ \omega^2 + D^2 k^4$ to $\omega^2 + v_s^2 k^2$ for
$k \lesssim \xi^{-1}$, with $v_s=\sqrt{n_s/\chi}$ the speed of sound
of the emerging LL and $\chi$ being the emerging finite compressibility \cite{MAX}. This changes the asymptotic
behavior of $V(\tau,x)$ to the conventional LL logarithmic growth at
scales longer than $\xi$, asymptotically resulting in the standard (1+1)-dimensional XY-model phenomenology.

\section{Irrelevance of Landau criterion}

In its original formulation, the Landau criterion of the stability of the superflow (a detailed discussion can be found in, e.g., Ref.~\cite{book}) relies on the system being Galilean invariant. Since our system is {\em not} Galilean invariant, it is instructive to see how the criterion changes when the system features an asymptotic---in the long-wavelength limit---translation invariance in the absence of Galilean invariance. We are concerned only with potential violation of the criterion in the infra-red limit. Therefore, it is sufficient to perform a general analysis at the level of the classical-field bilinearized hydrodynamic Hamiltonian, $H_{\rm hd}[\eta, \phi]$, written in terms of two canonically conjugate fields $\eta, \phi$. In the case of the LL, these are
small deviation  $\eta({\bf r})$ of the coarse-grained value of matter density $n$ from its equilibrium value $n_0$ and small deviation $\phi({\bf r})$ of the field of the superfluid phase from its equilibrium configuration  $\phi_0({\bf r}) = m \,  {\bf v}_0\cdot {\bf r}$. Here ${\bf v}_0$ is the velocity of the stationary superflow and $m$ is the bare atomic mass.

In the absence of the Galilean invariance, the ground-state energy density of the system, ${\cal E}(n, v_s)$, is a function of the density, $n$, and the magnitude of the superfluid velocity, $v_s$, with the following relations playing key hydrodynamic role \cite{book}:
\begin{equation}
 {\partial {\cal E}(n, v_s) \over \partial n }  \, =\, \mu\, , \qquad \quad {\partial {\cal E}(n, v_s) \over \partial v_s } \, =\,m j \, . \quad
 \label{Hydro_relations}
\end{equation}
Here  $\mu$ is the chemical potential, and $j$ is the magnitude of the supercurrent density, which is related to $v_s$ as $j= n_s v_s$, with $n_s$ the superfluid stiffness.

Performing an expansion of $\cal E$ around $n=n_0, v_s=v_0$, the Hamiltonian density ${\cal H}_{\rm hd}[\eta, \phi]$ follows---up to the lowest-order in $(\nabla \phi, \eta)$---as
\begin{equation}
{\cal H}_{\rm hd}(\eta, \phi) \, =\,   {\tilde{n}_s \over 2m } \, (\nabla \phi)^2 \, +\, {\eta^2 \over 2\chi} \,
\,+\, \lambda_s \eta \, {\bf v}_0 \cdot \nabla \phi \,  ,
\label{H_hydro}
\end{equation}
where
\begin{equation}
\tilde{n}_s\, =\,  {1\over m}\frac{\partial^2\cal E}{\partial v_0^2}\, =\, \frac{\partial j}{\partial v_0}\, =\, n_s  +v_0 \frac{\partial n_s(n_0,v_0)}{\partial v_0}\, ,
\label{tilde_n_s}
\end{equation}
and
\begin{equation}
\chi \, =\,  {\partial n(\mu, v_0) \over \partial \mu}\, , \qquad \quad   \lambda_s \, =\, {\partial n_s(n_0,v_0)  \over \partial n_0 } \, ,
 \label{partials}
\end{equation}
where $\chi$ the compressibility. Note that in the absence of the Galilean invariance, $n_s$, can be a nontrivial function of $(n, v_s)$.

The dispersion of the normal modes governed by the Hamiltonian (\ref{H_hydro}) is
\begin{equation}
\omega({\bf k}) \, =\, v_* k \, +\, \lambda_s {\bf v}_0\cdot {\bf k}\, , \qquad \quad v_*\, =\, \sqrt{\tilde{n}_s \over \chi} \, . \quad
 \label{disp}
\end{equation}
The critical value, $v_0^{(c)}$, of the magnitude of the superflow velocity corresponds to vanishing of $\omega({\bf k}) $ %for certain wavevectors.From (\ref{disp}) we see that
at the critical speed 
\begin{equation}
v_0^{(c)} \, =\, v_*/\lambda_s\, ,
 \label{Landau}
\end{equation}
so that $\omega({\bf k})$ can become negative for a range of $v_0 > v_0^{(c)} $.
In a Galilean invariant superfluid, $\tilde{n}_s=n_s = n$, implying $\lambda_s =1$ and  the critical velocity coincides with the velocity of sound: $v_0^{(c)}=v_*$. In a general case,
$\lambda_s$ can change not only its value but even the sign. For example, in a system of hard-core lattice bosons, $\lambda_s$ is close to unity at small fillings and approaches negative unity when the filling factor approaches unity. By continuity, this means that there is a special value of $n$ at which $\lambda_s=0$ and the generic Landau-type mechanism of instability becomes irrelevant. This does not exclude system-specific alternative mechanisms, but does imply that these
have do deal with essentially finite wavevectors.

In the  TQF the compressibility is infinite, $\chi =\infty$, and the role of the density $n$ is played by the dislocation climbing displacement $h$ (see Eq.(2) of the main text setting a relation between the dislocation displacement $\delta h$ and the density $\delta n$ variations), with the potential-energy density determined by the elastic deformation of the dislocation $\kappa^{-1} (\pt_x h)^2/2$. Furthermore, due to the translational invariance, $n_s$ can only depend on the curvature $\sim \pt^2_{xx} h$ of the dislocation and not on the  position $\sim n$ of the line \cite{Kuklov2022}. Accordingly, the spectrum $\omega \sim k^2$ at finite $v_0$ acquires the term $\sim v_0 k^3$. In the $k\to 0$ limit,  this correction becomes irrelevant (equivalently, we can put it as $\lambda_s\equiv 0$), and there is no room for the Landau-type
hydrodynamic instability.

\
\section{I-V characteristic}
  \begin{figure}[!htb]
%\vskip-8mm
	\includegraphics[width=0.9 \columnwidth]{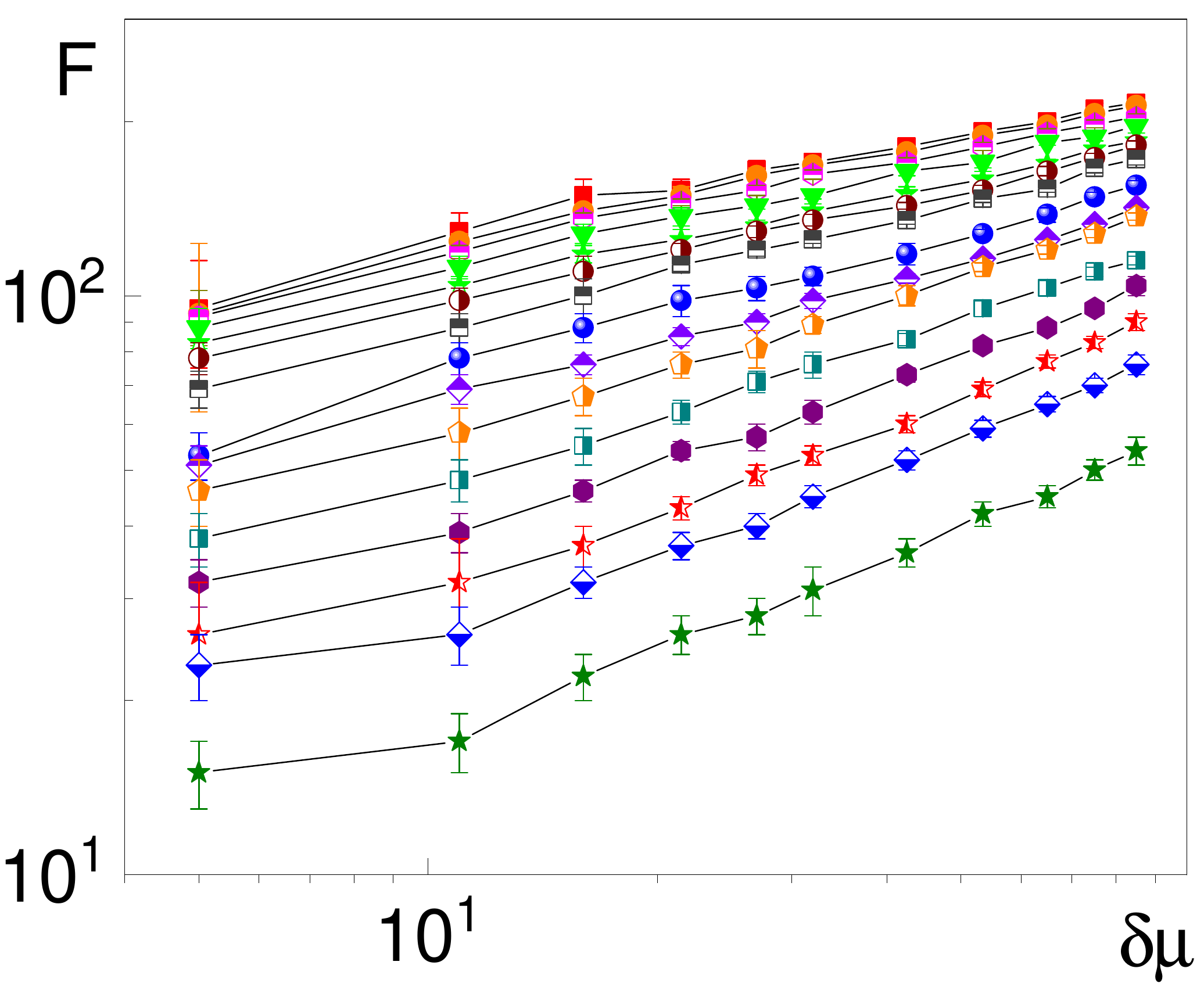}
%	\vskip-8mm
	\caption{(Color online) Raw data from Ref.~\cite{Moses} for the flux dependence on the chemical potential difference (in arbitrary units) at a range of temperatures. The lines are guides to an eye. Symbols are the same as in Fig. 2 of the main text and correspond to the same temperatures. }
	\label{fig2}
\end{figure}
  \begin{figure}[!htb]
%\vskip-8mm
	\includegraphics[width=1.0 \columnwidth]{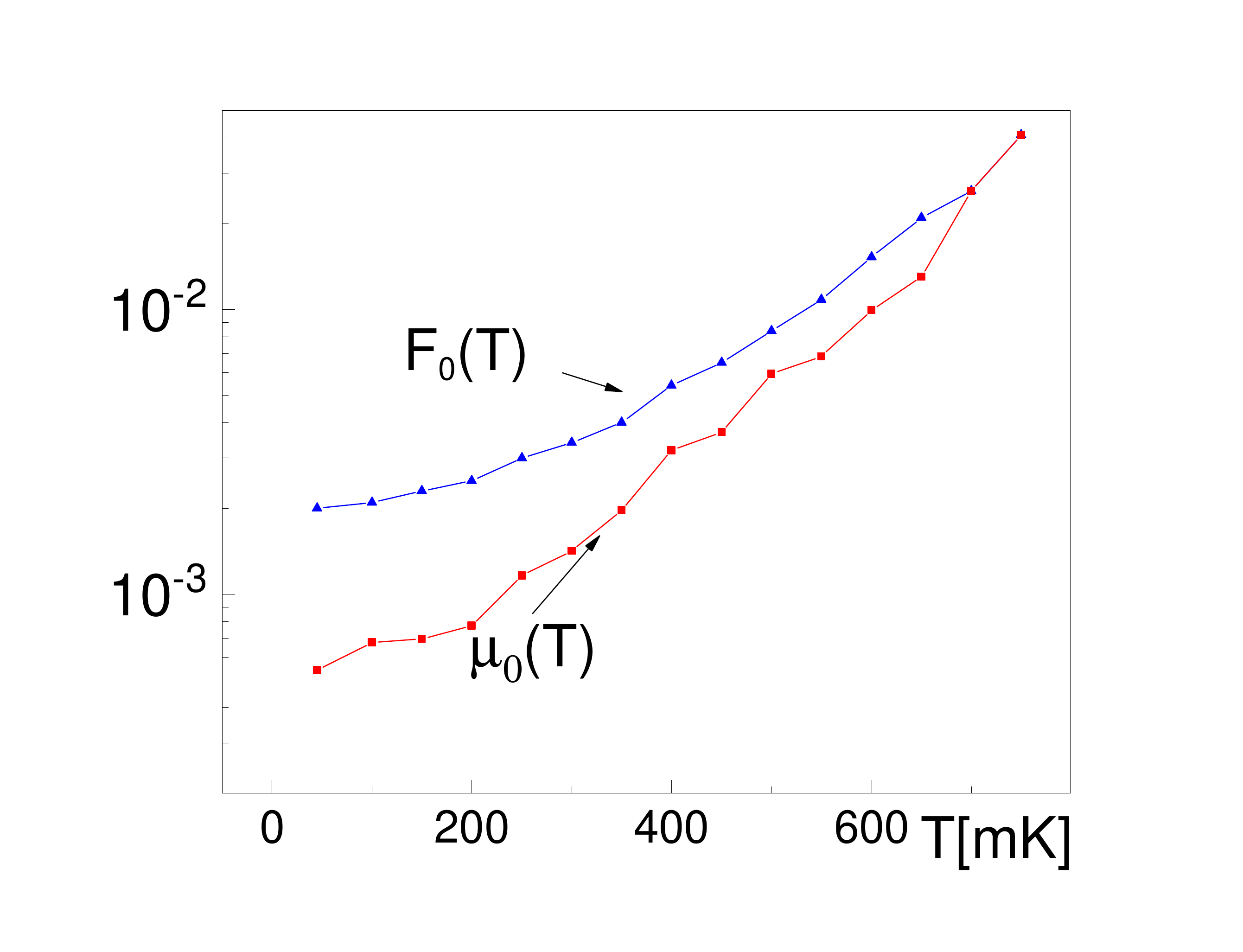}
%	\vskip-8mm
	\caption{(Color online)
Scaling factors $F_0(T)$ and $\mu_0(T)$ (symbols) used
in processing the data from Fig.5 of Ref.\cite{Moses}. Solid lines are guides to the eye. }
	\label{FM}
\end{figure}
The instanton mechanism of the superflow relaxation by phase slips is described here
in the leading exponential approximation when  $S^*_{\rm inst} >>1$, see Eq.(15) of the main text.
Without knowing the nature of the pre-exponential factor in the probability  (16) (in the main text), we use the Newtonian  approach in deriving the I-V characteristics -- Eq. (18) in the main text.  Phase slip events conclude with a jet of normal excitations carrying away
the superflow momentum---they produce a friction force $f=\tau^{-1} 2\pi n_s$
with the relaxation rate $\tau ^{-1} \propto  {\cal P}$, where ${\cal P}$ is given in Eq.(16) of the main text.
This force balances the force $f_{\rm bias} \approx n_s\Delta \mu $ imposed  by the bias,
and the stationary flow corresponds to $\Delta \mu = \tau^{-1} 2\pi $.

In order to compare this model with the experiment, an assumption should be made
about the proportionality coefficient in the relation $\tau ^{-1} \propto  {\cal P}$ as a function of the superflow velocity $v_0$.
We will follow the model of liquid friction, that is, $\tau^{-1} = (v_0/l_0) {\cal P}\equiv (v_0/l_0) \exp(-v_c/|v_0|)$, where $l_0$ is some length scale $l_0$ independent of $v_0$. This form respects the reflection symmetry $\Delta \mu \to -\Delta \mu$ and $v_0 \to - v_0$ together with the breaking of the time reversal due to the dissipation caused by the phase slips. Thus, the stationary flow velocity obeys
\be
(v_0/l_0) {\rm e}^{-v_c/|v_0|} \approx \Delta \mu .
\label{f}
\ee
In the dimensionless units ${\hat v}=v_0/v_c $ and $\mu=\delta \mu l_0/v_c$ this leads to
Eq.~(18) of the main text, which plays the role of universal master curve
presented in Fig 2 of the main text.
Its parts over about two orders of magnitude in $\mu$ and a factor of 4 in $\hat v$ can be approximated by the power law (see Eq.(17) of the main text) -- similarly to the experiments \cite{Hallock,Moses}.
One can define the effective exponent $\alpha $ using the log-derivative of the master curve, Eq.(18), of the main text:  $\alpha^{-1} = d \ln \mu /d \ln {\hat v}$, or, $\alpha = {\hat v}/(1+ {\hat v})$. By comparing this value with the
exponent observed for each data set taken at temperature $T$, see Figs.5(a) and (b) of Ref.~\cite{Moses} (presented above in Fig.~\ref{fig2} in arbitrary units ) , we determine the rescaling parameters $F_0(T), \mu_0(T)$ used to plot the experimental data together with
the master curve, Fig. 2 of the main text.
More specifically, each data set in the temperature range from $T=0.045$K to $T=0.75$K (characterized by the corresponding value of $\alpha(T)$) from Fig.5 of Ref.\cite{Moses} has been rescaled as $F \to F_0(T) F$ and $\Delta \mu \to \mu_0(T) \Delta \mu$ by some parameters in order to achieve the best fit by the dependence (18) of the main text. On the log-log graph, Fig.2 of the main text, this implies that the data at a given $T$ was shifted vertically and horizontally (without changing its slope).

Temperature dependence of the scaling factors is shown in Fig.~\ref{FM}.
It can be interpreted in terms of exponential suppression of the superfluid density $n_s$ by rare thermal fluctuations of the dislocation shape \cite{Kuklov2022}. This mechanism leads to the
stretched exponential dependence $\ln n_s(T) \propto - T^{5/4}$ on temperature
with weak links providing nucleation centers for efficient
phase slips. Since the instanton action is proportional to $\sqrt{n_s}$ (see Eq.(16) of the main text), this results in the growth of $\mu_0(T)$ and $\alpha(T)$ with increasing $T$.% is qualitatively consistent with the mechanism \cite{Kuklov2022}.


\begin{thebibliography}{99}

\bibitem{Hallock} M. W. Ray and R. B. Hallock, Phys. Rev. Lett. {\bf 100}, 235301
(2008); Phys. Rev. {\bf B 79}, 224302 (2009); M. W. Ray and R. B. Hallock, Phys. Rev. {\bf B 81}, 214523 (2010).

\bibitem{Hallock2012}    Ye. Vekhov and R. B. Hallock, Phys. Rev. Lett. {\bf 109}, 045303 (2012).

\bibitem{Hallock2019} R.B. Hallock, J. Low Temp. Phys. {\bf 197}, 167 (2019)

\bibitem{Beamish} Z. G. Cheng, J. Beamish, A. D. Fefferman, F. Souris, S. Balibar, and V. Dauvois,
Phys. Rev. Lett. {\bf 114}, 165301 (2015);  Z. G. Cheng and J. Beamish, Phys. Rev. Lett. {\bf 117}, 025301 (2016).

\bibitem{Moses} J. Shin, D. Y. Kim, A. Haziot, and M. H. W. Chan, Phys. Rev. Lett. {\bf 118}, 235301 (2017).

\bibitem{Moses2019} J. Shin and M. H. W. Chan, Phys. Rev. {\bf B 99}, 140502(R) (2019).

\bibitem{Moses2020} J. Shin and M. H. W. Chan, Phys. Rev. B {\bf 101}, 014507 (2020).


\bibitem{Moses2021} M.H.W. Chan, J. Low Temp. Phys. {\bf 205}, 235 (2021).


\bibitem{sclimb} S. G. S\"oyler, A. B. Kuklov, L. Pollet, N. V. Prokof'ev, and B. V. Svistunov, Phys. Rev. Lett. {\bf 103}, 175301(2009).

\bibitem{Kuklov2022} A. B. Kuklov, L. Pollet, N. V. Prokof'ev, and B. V. Svistunov, Phys. Rev. Lett. {\bf 128}, 255301 (2022).

\bibitem{SM} Supplemental material.

\bibitem{ColemanInstantons} S. Coleman, {\it Aspects of Symmetry}
  (CUP, NY, 1985).


\bibitem{ODLRO} L. Liu and  A. B. Kuklov, Phys. Rev. {\bf B 97}, 104510 (2018).

\bibitem{MAX} M. Yarmolinsky and A. B. Kuklov,  Phys. Rev. {\bf B 96},
  024505 (2017).

\bibitem{TonerNelson81} J. Toner and D. R. Nelson, Phys. Rev. B {\bf 23}, 316 (1981).

\bibitem{QSmGaugeLR} L. Radzihovsky, Phys. Rev. Lett. {\bf 125}, 267601 (2020); Z. Zhai,
  L. Radzihovsky,  Annals of Physics {\bf 435}, 168509 (2021).

\bibitem{book} B. Svistunov, E. Babaev, and N. Prokof'ev,  {\it Superfluid States of Matter}, Taylor \& Francis,  2015.





\end{thebibliography}
\end{document}